\documentclass[preprint,3p,times,twocolumn,sort&compress]{elsarticle}
\usepackage{graphicx}
\usepackage{epstopdf}
\usepackage{epsfig}
\usepackage{amsmath}
\usepackage{color}
\usepackage{slashed}
\usepackage{bm}
\usepackage{hyperref}

\newcommand{\al}{\alpha}
\newcommand{\be}{\beta}
\newcommand{\ga}{\gamma}
\newcommand{\Ga}{\Gamma}
\newcommand{\de}{\delta}

\newcommand{\ep}{\varepsilon}

\newcommand{\la}{\lambda}
\newcommand{\La}{\Lambda}
\newcommand{\si}{\sigma}

\renewcommand{\th}{\theta}   

\newcommand{\Om}{\Omega}

\DeclareMathOperator{\sech}{sech}

\newcommand{\beq}{\begin{align}}
\newcommand{\eeq}{\end{align}}
\newcommand{\ba}{\begin{array}}
\newcommand{\ea}{\end{array}}
\newcommand{\bea}{\begin{eqnarray}}
\newcommand{\eea}{\end{eqnarray}}
\newcommand{\bi}{\begin{itemize}}
\newcommand{\ei}{\end{itemize}}
\newcommand{\ben}{\begin{enumerate}}
\newcommand{\een}{\end{enumerate}}
\newcommand{\bc}{\begin{center}}
\newcommand{\ec}{\end{center}}
\newcommand{\bl}{\begin{flushleft}}
\newcommand{\el}{\end{flushleft}}
\newcommand{\br}{\begin{flushright}}
\newcommand{\er}{\end{flushright}}

\newcommand\Eqn[1]{Eq.~(\ref{#1})}  
\newcommand{\mr}{\mathrm}
\newcommand{\mb}{\mathbf}

\newcommand{\mc}{\mathcal}
\newcommand{\mi}{\mathop{}\!i}
\newcommand{\me}{\mathop{}\!e}
\newcommand{\dif}{\mathop{}\!d}
\newcommand{\p}{\partial}

\newcommand{\tr}{\hbox{tr}}
\newcommand{\Tr}{\hbox{Tr}}

\newcommand{\<}{\langle}
\renewcommand{\>}{\rangle}   
\renewcommand{\l}{\left}
\renewcommand{\r}{\right}

\newcommand\comment[1]{ \hbox{[{\it Comment suppressed here.}\/]} }
\newcommand\hide[1]{}
\newcommand{\skipover}[1]{}

\begin{document}

\begin{frontmatter}

\title{Charged pion condensation under parallel electromagnetic fields}
\author[l1]{Jingyi Chao}
\ead{jychao@impcas.ac.cn}
\author[l2,l3]{Mei Huang}
\ead{huangm@mail.ihep.ac.cn}
\author[l1,l4]{Andrey Radzhabov}
\ead{aradzh@icc.ru}
\address[l1]{Institute of Modern Physics, Chinese Academy of Sciences, Lanzhou, 730000, China}
\address[l2]{Institute of High Energy Physics, Chinese Academy of Sciences, Beijing, 100049, China}
\address[l3]{School of Physics Sciences, University of Chinese Academy of Sciences, Beijing 100039, China}
\address[l4]{Matrosov Institute for System Dynamics and Control Theory
 , 664033, Irkutsk, Russia}

\author{}

\address{}

\begin{abstract}

The formation of charged pion condensate under parallel electromagnetic fields is studied within the two-flavor Nambu--Jona-Lasinio model. The technique of Schwinger proper time method is extended to explore the quantity locating in the off-diagonal flavor space, i.e., charged pion. We obtain the associated effective potential as a function of the strength of the electromagnetic fields and find out that it contains a sextic term which possibly induce weakly first order phase transition. Dependence of pion condensation on model parameters is investigated.

\end{abstract}

\end{frontmatter}

\section{Introduction}

The phase structure of Quantum Chromodynamics (QCD) at high temperature/density and other extreme conditions has attracted lots of attentions and been a main topic of heavy ion collisions. The perturbative QCD predicts a free gas of quarks and gluons at high temperature limit and a color-flavor-locking phase at very high baryon density but low temperature. However, the QCD vacuum has a rather complicated nonperturbative structure, and the QCD phase diagram is not a simple transition between the hadron phase with non-zero chiral condensate to the weakly coupled quark-gluon plasma as expected long time ago \cite{Cleymans:1985wb}, but instead a rich structure of different phases with corresponding condensates. These phases could include different color superconducting states or inhomogeneous chiral condensates \cite{Buballa:2003qv,Buballa:2014tba,Andersen:2018osr}. Recently, QCD phase structure under strong magnetic fields has drawn great interests ~\cite{Kharzeev:2007jp,Skokov:2009qp,Hattori:2016emy,Andersen:2014xxa,Miransky:2015ava,Huang:2015oca}. The strong magnetic fields can be generated with the strength up to $B\sim10^{18\sim 20}$ G in the non-central heavy ion collisions \cite{Skokov:2009qp,Deng:2012pc}, and
is expected to be on the order of $10^{18}$-$10^{20}$ G~\cite{Duncan:1992hi,Blaschke:2018mqw} in the inner core of magnetars.

Lots of interesting phenomena under strong magnetic fields have been discussed, for example, the magnetic catalysis~\cite{Klevansky:1989vi,Klimenko:1990rh,Gusynin:1995nb,Gusynin:1999pq}, inverse magnetic catalysis~\cite{Bali:20111213,Bali:2012zg,Bali:2013esa} effect, the chiral magnetic effect (CME)~\cite{Kharzeev:2007jp,Kharzeev:2007tn,Fukushima:2008xe} and the vacuum superconductivity~\cite{Chernodub:2010qx,Chernodub:2011mc}.
Moreover, it was pointed out that under the parallel electromagnetic fields, the neutral pion condensation can be formed~\cite{Cao:2015cka,Wang:2018gmj}
due to the connection of field with axial anomaly. If only QCD interaction is included, the axial isospin currents is anomaly free. It turns out that anomaly emerges associated with the coupling of quarks to electromagnetism, where the axial isospin currents is given by
\begin{equation}
	\p_{\mu}j_{5}^{\,\mu 3}=-\frac{e^{2}}{16\pi^{2}}\ep^{\al\be\mu\nu}F_{\al\be}F_{\mu\nu}\cdot\tr\l[\tau^{3}Q^{2}\r].
\end{equation}
Here $Q$ is the matrix of quark electric charges and $F$ is the field strength. The corresponding process is $\pi_{0}\to \ga\ga$. The decay of a neutral pion into two photons, which had been a puzzle for some time in the 1960s, is the most successful proof of chiral anomaly. Above solution led to the discovery of the Adler--Bell--Jackiw anomaly \cite{Adler:2004qt}.

In the asymmetric flavor space, one can introduce a chiral isospin chemical potential $\mu_{I}^5$ corresponding to the current $\bar{\psi}\ga_{0}\ga_{5}\tau_{3}\psi$, which is similar to the isospin chemical potential $\mu_{I}$ with respect to $\bar{\psi}\ga_{0}\tau_{3}\psi$.
It has been a long history of investigating the pion condensation under the isospin asymmetric nuclear matter.
In the beginning this effect is discussed for case nuclear matter in neutron-star interiors \cite{Sawyer:1972cq,Sawyer:1973fv,Voskresensky:1980nk} or superdense and supercharged nuclei \cite{Migdal:1978az}. The pion condensation of charged or neutral pion modes in QCD vacuum are also considered in the frameworks of effective models with quark degrees of freedom \cite{Son:2000xc,He:2006tn,Mao:2014hga,Khunjua:2017khh} or in lattice calculations \cite{Brandt:2016zdy,Brandt:2018bwq}.

The degeneracy between $\pi_{0}$ and $\pi_{\pm}$ is destroyed because of the axial isospin chemical potential. It is worth to pursuing the detailed behaviors of charged pions in a strict manner. Hence, in this work, we focus on the possibility of charged pion condensation under the parallel electromagnetic fields in the framework of the $\mr{SU}(2) \times \mr{SU}(2)$ NJL model \cite{Nambu:1961tp,Nambu:1961fr}. For this purpose, we develop a full routine to derive the mean-field thermodynamical potential of the NJL model with nonzero charged pion condensate $\< \bar{\psi} i \gamma_5 \tau_{\pm} \psi \>$  in the off-diagonal flavor space under the parallel electromagnetic fields. Calculations are performed with Schwinger proper time method \cite{Schwinger:1951nm} and the proper time regularization in the NJL model is used.
Through the paper we only consider the model at zero temperature and chemical potential and restrict ourselves to the case of the electric field anti-parallel to the magnetic field.

\label{intr}
\section{Lagrangian}\label{NJL}
The Lagrangian of the $\mr{SU}(2)\times \mr{SU}(2)$ NJL model is in the form of \cite{Nambu:1961tp,Nambu:1961fr,Volkov:1986zb,Vogl:1991qt,Klevansky:1992qe,Hatsuda:1994pi,Volkov:2005kw}
\begin{align}
\mathcal{L}_{\mr{NJL}} = \bar{\psi}\left(i \slashed{D} - m_0 \right) \psi
+ \mathrm{G} \left[ \left(  \bar{\psi} \psi \right)^2+ \left(  \bar{\psi} i\gamma_5 \tau_i \psi \right)^2 \right],
\end{align}
where $\bar{\psi}(x)=(\bar{u}(x),\bar{d}(x))$ are  $u$ and $d$ anti-quark fields. The limit of equal current masses for $u,d$,  $m_u=m_d\equiv m_{0}$ is considered. $\gamma_i$, $\tau_{i}$ are conventional Dirac and Pauli matrices and $\tau_{0}$ is the unit matrix.
$\slashed{D}$ is the covariant derivative and, in the two flavor space, expressed as
\begin{align}
	D_{\mu}=\l(\p_{\mu}-\mi QA_{\mu}\r)\tau_{0}-\mi qA_{\mu}\tau_{3},
\end{align}
where $Q=\frac{1}{2}\l(q_{u}+q_{d}\r)$ and $q=\frac{1}{2}\l(q_{u}-q_{d}\r)$.

Introducing auxiliary bosonic fields $\pi$, $\sigma$, with the help of Hubbard-Stratonovich transformation one can integrate over the quark fields,
then obtains the following effective Lagrangian:
\begin{align}
\mathcal{L} = \frac{\sigma^2+\vec{\pi}^2}{4\mathrm{G}}-\mi \Tr \ln S^{-1}, \label{eLagrangian}
\end{align}
where $S^{-1}$ is the inverse quark propagator and
\begin{align}
S^{-1}= \mi\slashed{D} - M, \quad M=m_{0}\tau_{0}-\sigma\tau_{0}-\mi \gamma_5\pi_{i} \tau_i. \label{Propagator}
\end{align}
The auxiliary bosonic fields could have a nonzero vacuum expectation values and therefore it is necessary to shift them as $\sigma = \sigma^\prime-\<\sigma\>$, $\pi_{i}=\pi_{i}^\prime- \<\pi_{i}\>$. 
Equations of motion for mean-fields $\<\sigma\>$, $\<\pi_{i}\>$ are obtained from the Lagrangian (\ref{eLagrangian}) after
elimination from its linear terms, i.e.
\begin{align}
\frac{\delta \mathcal{L} }{\delta \<\sigma\>}\biggl|_{\substack{ \sigma^\prime =0 \\ \pi_{i}^\prime=0 }}, \quad
\frac{\delta \mathcal{L} }{\delta \<\pi_{i}\>}\biggl|_{\substack{ \sigma^\prime =0 \\ \pi_{i}^\prime=0 }} =0.
\end{align}
As a result, under different conditions the $\<\si\>$, $\<\pi_{i}\>$ condensates have non-zero values and the non-zero value of scalar condensate leads to a formation of constituent quarks with dynamical quark mass $m=m_{0}-\<\si\>$.

Let us denote the second term of effective Lagrangian (\ref{eLagrangian})  as $\mathcal{S}_{eff}=-\mi \Tr \ln S^{-1}$.
Then the gap equations for $\<\si\>$ and $\<\pi_{i}\>$  takes the form
\begin{align}
m=m_{0}-2\mathrm{G}\frac{\partial\mathcal{S}_{eff}}{\partial \<\sigma\>},\quad
\<\pi_{i}\>=-2\mathrm{G}\frac{\partial \mathcal{S}_{eff}}{\partial \<\pi_{i}\>}.
\end{align}

The calculation of $\mathcal{S}_{eff}$ is presented in the following section.

\section{The effective potential}\label{main}

Without loss of generality, one can choose $\<\pi_{i}\>=\l(\pi_{1},0,0\r)$ and therefore ``mass'' in quark propagator Eq.(\ref{Propagator}) is $M=m\tau_{0}+\mi\pi_{1}\ga_{5}\tau_{1}$.
Since $\mr{Det}\l(\mi\slashed{D}-M\r)=\mr{Det}\,\Ga\l(\mi\slashed{D}-M\r)\Ga$, where $\Ga=\ga_{5}\tau_{3}$, the second term of the Lagrangian Eq.(\ref{eLagrangian}) is replaced to
\begin{equation}
	\mathcal{S}_{eff}=-\frac{\mi}{2}\ln\mr{Det}\l(\slashed{\mathcal{D}}^{2}+m^{2}+\pi_{1}^{2}\r),
\end{equation}
where $\slashed{\mathcal{D}}^{2}=\slashed{D}^{2}-\ga_{5}\ga^{\mu}\pi_{1}\l[\tau_{1},D_{\mu}\r]$.

By using the method of proper time, we represent $\mathcal{S}_{eff}$ as following:
\begin{equation}\label{eqn_seff}
	\mathcal{S}_{eff}=\Tr\int\limits_{1/\La^{2}}^{\infty}\mi\,\frac{\dif s}{2s}\int\tr\l\<x\big|\me^{-\mi\l(\slashed{\mathcal{D}}^{2}+m^{2}+\pi_{1}^{2}\r)s}\big|x'\r\>\dif^{4}x,
\end{equation}
where the ultraviolet cutoff $1/\La^{2}$ has been explicitly introduced, $\tr$ and $\Tr$ means the trace taking in the spinor and flavor space, respectively.

From now on, we will work in the Euclidean space. Following notations are introduced:
\begin{align}\label{AlBeLa}
&	\al=m^{2}+\pi_{1}^{2}-\frac{1}{2}\si^{\mu\nu}\la_{\mu\nu},\quad
	\be_{\nu}=q\pi_{1}\ga_{5}\ga^{\mu}F_{\mu\nu}\tau_{2},\quad\nonumber \\
&    \la_{\mu\nu}=q_{f}F_{\mu\nu},
\end{align}
where $q_{f}=\mr{Diag}(q_{u},q_{d})$ and $\si^{\mu\nu}=\frac{\mi}{2}\l[\ga^{\mu},\ga^{\nu}\r]$.
In order to obtain $\mathcal{S}_{eff}$, it is then straightforward to look for the solution of $G(x,y;s)$ 
obeying a second order differential equation $\l(\slashed{\mathcal{D}}^{2}+m^{2}+\pi_{1}^{2}\r)G\l(x,y;s\r)=\de\l(x,y;s\r)$. 
The explicit form is
\begin{align}
\slashed{\mathcal{D}}^{2}+m^{2}+\pi_{1}^{2}&=\p^{2}_{x}+\alpha(y)+\beta_{\mu}(y)\l(x-y\r)^{\mu}+\nonumber\\
&	+\frac{1}{4}\la^{2}_{\mu\nu}\l(x-y\r)^{\mu}\l(x-y\r)^{\nu}.
\end{align}
Performing the Fourier transform, one finds,
\begin{align}\label{eqn_diff_p}
	\l(-p^{2}+\alpha-\mi\beta_{\mu}\frac{\p}{\p p_{\mu}}-\frac{1}{4}\la^{2}_{\mu\nu}\frac{\p^{2}}{\p p_{\mu}\p p_{\nu}}\r)G(p;s)=1.
\end{align}
As suggested in the the reference \cite{Brown:1975bc} 
one can solve the equation in the form 
\begin{align}\label{eqn_G_pV2}
	G(p;s)=\me^{-\alpha s}\me^{ p\cdot A(s)\cdot p+B(s) \cdot p+C(s)},
\end{align}
whose associated descriptions of matrix $A$, vector $B$ and scalar $C$ are
\begin{align}\label{eqn_sol_ABC}
&A=\la^{-1}\tan\la s,\quad
B=-2\mi\be\cdot\la^{-2}\l(1-\sec\la s\r),\\
&C=-\frac{1}{2}\,\tr\ln\cos\la s-\be\cdot\la^{-3}\l(\tan\la s-\la s\r)\cdot\be \nonumber.
\end{align}
For simplicity here and below indexes are not shown.

Plugging the form of $\be$ in Eq.(\ref{AlBeLa}) into vector $B$ and restoring indexes one has
\begin{align}
	B_{\mu}=-2\mi q\pi_{1}\tau_{2}\ga_{5}\ga^{\nu}F_{\nu\al}\l[\la^{-2}\l(1-\sec\la s\r)\r]^{\al}_{\mu}.
\end{align}
Vector $B$ contains Dirac matrix, not commuting with $\si^{\mu\nu}$.
Therefore, we emphasize that one should be careful with tracing in spinor space and integrating in momentum space.

Introducing notations $P_{1}=\frac{1}{2}\si\la s$ and $P_{2}=p\cdot A(s)\cdot p+B(s) \cdot p$, one has $\l[\si\la s, p\cdot A(s)\cdot p\r]=0$ and
the part with matrices in exponent Eq.(\ref{eqn_G_pV2}) can be expanded as
\begin{align}
&\me^{P_{1}+P_{2}}\simeq \me^{P_{1}}\me^{P_{2}}\me^{-\frac{1}{2}[P_{1},P_{2}]}=\nonumber \\
&\quad\quad=\me^{\frac{1}{2}\si\la s}\me^{p\cdot A(s)\cdot p+B(s) \cdot p}\me^{-\frac{1}{4}\l[\si\la s,B(s) \cdot p\r]}.
\end{align}
We denote $-\frac{1}{4}\l[\si\la s,B(s) \cdot p\r]=\frac{1}{2}q\pi_{1} Os$, where $O$ has a structure of the form  $O=Q\tau_{2}O_{1}\mathbb{B}_{1}p+q\tau_{1}O_{2}\mathbb{B}_{2}p$ and $\mathbb{B}$ will render in Eq. (\ref{Bdefinition}). Shorthand matrix notation is applied, i.e. $\mathbb{F}=F_{\mu}^{\nu}$. To find the eigenvalue of $O$, we square it and get
\begin{align}
O^{2}&=Q^{2}\l(\tau_{2}O_{1}\mathbb{B}_{1}p\r)^{2}+q^{2}\l(\tau_{1}O_{2}\mathbb{B}_{2}p\r)^{2} -\nonumber \\
&-\mi qQ\tau_{3}\l[O_{1}\mathbb{B}_{1}p,O_{2}\tilde{\mathbb{B}}_{2}p\r],\\
&O_{1}=\mi\l[\si_{\mu\nu},\ga_{5}\ga^{\al} \r]=2\ga_{5}g_{\nu}^{\al}\ga_{\mu}-2\ga_{5}g_{\mu}^{\al}\ga_{\nu},\nonumber\\
&O_{2}=\l\{\si_{\mu\nu},\ga_{5}\ga^{\al}\r\}=-2\ep^{\al\be}_{\;\mu\nu}\ga_{\be}.\nonumber
\end{align}
With help of relation $\tau_{2}q_{f}\tau_{2}=\mr{Diag}\l(q_{d}, q_{u}\r)=\tilde{q}_{f}$, the $\tilde{\mathbb{B}}\l(\mathbb{B}\r)$ are shown as
\begin{align}\label{Bdefinition}
&\tilde{\mathbb{B}}_{1}\l(\mathbb{B}_{1}\r)=\frac{1}{\mathsf{q}^{2}}\l[1-\sec \mathsf{q}\mathbb{F}s\r],\quad \nonumber \\
&\tilde{\mathbb{B}}_{2}\l(\mathbb{B}_{2}\r)=
\frac{\bar{\mathbb{F}}\mathbb{F}}{\mathbb{F}^{2}}
\frac{1}{\mathsf{q}^{2}}
\l[1-\sec\mathsf{q}\mathbb{F}s\r],
\end{align}
where $\mathsf{q}=\tilde{q}_{f}$ or $q_{f}$ for $\tilde{\mathbb{B}},\mathbb{B}$ respectively; $\mathbb{F}$ and $\bar{\mathbb{F}}$ are field strength tensor $F^{\mu\nu}$ and dual field strength tensor $\bar{F}^{\mu\nu}=\frac{1}{2}\ep^{\mu\nu\al\be}F_{\al\be}$, in shorthand notations.   Moreover, $\l(\tau_{2}O_{1}\mathbb{B}_{1}p\r)^{2}=-16\mathbb{B}_{1}\tilde{\mathbb{B}}_{1}p^{2}$, $\l(\tau_{1}O_{2}\mathbb{B}_{2}p\r)^{2}=16\mathbb{B}_{2}\tilde{\mathbb{B}}_{2}p^{2}$ and $[O_{1}\mathbb{B}_{1}p,O_{2}\tilde{\mathbb{B}}_{2}p]=-32\ga_{5}\mathbb{B}_{1}\tilde{\mathbb{B}}_{2}p^{2}$.

Applying the system that in a Lorentz frame where the electromagnetic field vectors are anti-parallel, e.g., $\mb{B}=-\mb{E}=f\hat{z}$\,, one gets $\mathbb{F}^{2}=f^2\,\mr{Diag}\l(-,+,+,-\r)$ and $\bar{\mathbb{F}}\mathbb{F}=-f^2\de_{\mu\nu}$ in Euclidean metric $(-,-,-,-)$, hence that $\bar{\mathbb{F}}\mathbb{F}/\,\mathbb{F}^{2}=f^{2}\mathbb{F}^{-2}$. Besides, $\l[1-\sec\mathsf{q}\mathbb{F}s\r]$ contains even powers of $\mathbb{F}$. It causes $O^{2}=-16Q^{2}\mathsf{p}_{1}^{2}+16q^{2}\mathsf{p}_{2}^{2}+32\mi\ga_{5}Qq\mathsf{p}_{1}\cdot\mathsf{p}_{2}$ in a simply manner, where $\mathsf{p}_{1}=p_{\shortparallel}+p_{\perp}$, $\mathsf{p}_{2}=p_{\shortparallel}-p_{\perp}$, $p_{\shortparallel}=b_{\shortparallel}(p_{0},0,0,p_{3})$ and $p_{\perp}=b_{\perp}(0,p_{1},p_{2},0)$. The forms of $b_{\shortparallel}$ and $b_{\perp}$ are taken as
\begin{equation}\label{eqn_b03}
	b_{\shortparallel}=\frac{(1-\sec q_{f}s)^{\frac{1}{2}}(1-\sec\tilde{q}_{f}s)^{\frac{1}{2}}}{q_{f}\tilde{q}_{f}},
\end{equation}
\begin{equation}\label{eqn_b12}
	b_{\perp}=\frac{(1-\sech q_{f}s)^{\frac{1}{2}}(1-\sech\tilde{q}_{f}s)^{\frac{1}{2}}}{q_{f}\tilde{q}_{f}}.
\end{equation}
Here and below in  we rescale the integration variable as $s=s^\prime/f$ and omit prime.
Because $\ga_{5}^{2}=1$ associated with eigenvalue $\pm 1$, it follows that $O$ has four eigenvalues \cite{Schwinger:1951nm}, written as
\begin{align}\label{eqn_O_squar}
	\mathcal{O}=\pm 4\l(\mi Q\mathsf{p}_{1}\pm q\tau_{3}\mathsf{p}_{2}\r).
\end{align}

Let $\th=q\pi_{1} s/f$, one has
\begin{align}
	\tr\,\me^{\frac{1}{2}\th O}=\mathsf{T}=\cos\l(2Q\th\mathsf{p}_{1}\r)\cosh\l(2\tau_{3}q\th\mathsf{p}_{2}\r),
\end{align}
which follow the method applied in \cite{dittrich2000probing}. The full statement is that 
\begin{align}\label{eqn_O_exp}
	\exp\l[\frac{1}{2}\th O\r]=\mathsf{T}+\mi\ga_{5}\mathsf{U}+\frac{O\mathsf{V}}{2K^{2}}+\frac{\mi\ga_{5}O\mathsf{W}}{2K^{2}},
\end{align}
where $K^{2}=\mathsf{p}_{1}^{2}=\mathsf{p}_{2}^{2}$. $\mathsf{T}, \mathsf{U}, \mathsf{V}$ and $\mathsf{W}$ are scalars. Similarly,
\begin{align}\label{eqn_F_exp}
	\exp\l[q_{f}\frac{\si Fs}{2f}\r]=\mathsf{P}-\mi\ga_{5}\mathsf{Q}+\frac{\si F}{2f}\,\mathsf{R}-\frac{\mi\ga_{5}\si F}{2f}\,\mathsf{S}.
\end{align}
Since
\begin{align}
	\tr\l(O^{2}\me^{\frac{1}{2}\th O}\r)=\frac{\p^{2}}{\p^{2}\th}\tr\l(4\me^{\frac{1}{2}\th O}\r)=4\frac{\p^{2}\mathsf{T}}{\p^{2}\th},
\end{align}
apply the identity of \Eqn{eqn_O_squar}, it derives that
\begin{align}
	\mathsf{U}=\sin\l(2Q\th\mathsf{p}_{1}\r)\sinh\l(2\tau_{3}q\th\mathsf{p}_{2}\r).
\end{align}
Proceeding with the direct differentiation of the exponential function via our basic trick, we get
\begin{align}
&\mathsf{V}=\frac{1}{Q^{2}+q^{2}}
\biggl(Q\mathsf{p}_{1}\sin\l(2Q\th\mathsf{p}_{1}\r)\cosh\l(2\tau_{3}q\th\mathsf{p}_{2}\r)+\nonumber\\
&\quad\quad\quad+\tau_{3}q\mathsf{p}_{2}\cos\l(2Q\th\mathsf{p}_{1}\r)\sinh\l(2\tau_{3}q\th\mathsf{p}_{2}\r)\biggr), \nonumber\\
&\mathsf{W}=\frac{1}{Q^{2}+q^{2}}
\biggl(\tau_{3}q\mathsf{p}_{2}\sin\l(2Q\th\mathsf{p}_{1}\r)\cosh\l(2\tau_{3}q\th\mathsf{p}_{2}\r)-\nonumber\\
&\quad\quad\quad-Q\mathsf{p}_{1}\cos\l(2Q\th\mathsf{p}_{1}\r)\sinh\l(2\tau_{3}q\th\mathsf{p}_{2}\r)\biggr).
\end{align}
Known in \cite{dittrich2000probing}, one has
\begin{align}
&\mathsf{P}=\cos q_{f}s\cosh q_{f}s,\, \mathsf{Q}=\sin q_{f}s\sinh q_{f}s \nonumber\\
&\mathsf{R}=(\sinh q_{f}s\cos q_{f}s+\cosh q_{f}s\sin q_{f}s)/2,\\
&\mathsf{S}=(\sinh q_{f}s\cos q_{f}s-\cosh q_{f}s\sin q_{f}s)/2.\nonumber
\end{align}
Then, we perform an approximate expansion
\begin{align}\label{eqn_ap2bp}
&\me^{ p\cdot A(s)\cdot p+B(s)\cdot p}\simeq
\me^{p\cdot A(s)\cdot p}\me^{B(s)\cdot p}
\\
&=\me^{ p\cdot A(s)\cdot p}\l(\cos\varrho
+B(s) \cdot p\frac{\sin\varrho}{\varrho}\r)\nonumber
 \end{align}
where $\varrho=2q\pi_{1}k/f$ and $k=\l(\mathsf{p}_{1}\cdot\mathsf{p}_{2}\r)^{\frac{1}{2}}$.

Now, it is allowed us to integrate with respect to $p$ and take the trace in the spinor space. With help of the \Eqn{eqn_O_exp}, \Eqn{eqn_F_exp} and \Eqn{eqn_ap2bp}, one has
\begin{align}
L(s)&=\tr\int\me^{\frac{1}{2f}\si\la s}\me^{p\cdot A(s)\cdot p+B(s)\cdot p}\me^{-\frac{1}{4f}\l[\si\la s,B(s) \cdot p\r]}\dif^{4}p \nonumber\\
&=L_{0}(s)+L_{1}(s)+L_{2}(s).
\end{align}
Here $\<X\>$ denotes integrating in momentum and tracing in spinor space $\tr\int X\me^{p\cdot A\cdot p}\dif^{4}p$. It gives
\begin{align}
&L_{0}(s)=\l\<\cos\varrho\mathsf{T}\mathsf{P}\r\>, \, \,
L_{1}(s)=\l\<\cos\varrho\mathsf{U}\mathsf{Q}\r\>, \, \, L_{2}(s)=\\
&=\l< \frac{2\tilde{q}_{f}\sin\varrho}{K^{2}k}
\l[q_{f}K^{2}\l(\mathsf{W}\mathsf{S}-\mathsf{V}\mathsf{R}\r)+\tilde{q}_{f}k^{2}\l(\mathsf{V}\mathsf{S}+\mathsf{W}\mathsf{R}\r)\r]\r>. \nonumber
\end{align}
The integration with respect to momentum $p$ is in the Gaussian form, which can be taken easily with result 
\begin{align}
&\<1\>=\mathcal{N}=\pi^{2}\mr{Det}A^{-\frac{1}{2}},\,A=\mr{Diag}\l(a_{\shortparallel}, a_{\perp}, a_{\perp}, a_{\shortparallel}\r)\nonumber\\
&\<K^{2}\>=\frac{\mathcal{N}}{2}\,\tr\l(\frac{D_{+}}{A}\r),\,
\<k^{2}\>=\frac{\mathcal{N}}{2}\,\tr\l(\frac{D_{-}}{A}\r)
\end{align}
The matrices $D_{\pm}=\mr{Diag}\l(b^{2}_{\shortparallel}, \pm b^{2}_{\perp}, \pm b^{2}_{\perp}, b^{2}_{\shortparallel}\r)$, which read from \Eqn{eqn_b03} and \Eqn{eqn_b12}. From \Eqn{eqn_sol_ABC}, one has
$a_{\shortparallel}=\tan q_{f}s/\l(q_{f} f\r)$ and $a_{\perp}=\tanh q_{f}s/\l(q_{f} f\r)$. The higher orders corrections $\<K^{4}\>$, $\<k^{4}\>$ and $ \<k^{2}K^{2}\>$ can be drawn in a similar manner, which are abbreviated here. 

Since $\th p\sim\pi_{1}ps/f\sim \pi_{1}p/\La^{2}\ll 1$ and the integration is exponential suppressed for large $s$, it enables us to approximate $\sin(a\th p)$, $\sinh(a\th p)\sim a\th p$ and $\cos(a\th p)$, $\cosh(a\th p)\sim 1$. Hence, it acquires $\mathsf{T}\sim 1$, $\mathsf{U}\sim k^{2}s$,  $\mathsf{V}\sim K^{2}s$ and $\mathsf{W}\sim  K^{2}k^2 s$. Finally, take the integration with respect to $s$ to get
\begin{align}\label{eqn_eff_potentioal}
&\mathcal{S}_{eff}= \mc{S}_{eff}^{0}+\mc{S}_{eff}^{1}+\mc{S}_{eff}^{2}, \\
&\mc{S}_{eff}^{i}=\frac{N_c }{4\pi^{2}}\Tr\int_{f/\La^{2}}^{\infty} \frac{\dif s}{2s} \me^{-h(s)} S_{eff}^{i}(s),\quad \nonumber 
\end{align}
where $-h(s)=-(m^{2}+\pi_{1}^{2})s/f+{C}(s)-\frac{1}{2}\ln\tr{A}$, and
\begin{align}
&{C}(s)-\frac{\ln\tr{A}}{2}=
-\ln\frac{\sin q_{f}s\sinh q_{f}s}{q_{f}^{2}f^{2}}\nonumber\\
&\quad\quad\quad-\frac{2q^{2}{\pi_{1}}^{2}}{\tilde{q}_{f}^{3}f}\l(2\tilde{q}_{f}s-\tan \tilde{q}_{f}s-\tanh \tilde{q}_{f}s\r).
\end{align}
The detailed integrands $S_{eff}^{i}(s)$ are 
\begin{align}
&S_{eff}^{0}(s) = \mathsf{P},\quad \nonumber \\
&S_{eff}^{1}(s) =4 \tau_{3} \frac{Q q^3\pi_{1}^2 s^2}{f^2\mc{N}}\frac{}{}\l\< k^2\r\> \mathsf{Q} ,\quad \label{exprS2} \\
&S_{eff}^{2}(s) =\frac{8\tilde{q}_{f}q^2\pi_{1}^2 s}{f^2\mc{N}}
\l(-q_{f}\l\< K^2\r\>\mathsf{R}+\tilde{q}_{f}\l\< k^2\r\>\mathsf{S}\r) 
+\nonumber\\
&\quad\quad+\tau_{3} \frac{32\tilde{q}_{f} Qq^5 \pi_{1}^4 s^3}{3f^4\mc{N}}
\l(q_{f}\l\< K^2k^2\r\>\mathsf{S}+\tilde{q}_{f}\l\< k^4\r\>\mathsf{R}\r).\nonumber  
\end{align}

Eventually, we have the effective potential which takes the following form:
\begin{align}
\Omega = \frac{(m-m_{0})^2+\pi_1^2}{4\mathrm{G}} + \mathcal{S}_{eff}.
\end{align}

\section{Numerical results}

The NJL model is nonrenormalizable and therefore the UV cut-off should be employed in order to get reasonable results, where a proper time regularization is applied in the work, i.e., 
the integration with respect to $s$ start from $f/\La^{2}$. 
We perform calculations of integral expression for $\mathcal{S}_{eff}$ in Eq. (\ref{eqn_eff_potentioal}) numerically.
In the limit of zero field $f$ the expression leads to the usual proper-time regularization scheme of NJL model.
Therefore, for numerical estimation we use the model parameterization from Ref. \cite{Inagaki:2015lma}.
Namely, in \cite{Inagaki:2015lma} there are five sets of model parameters for proper-time regularization scheme 
which are fitted in favor of observable values of pion mass and weak pion decay constant.
For convenience we present them in Table 1.
In the set 1 the constituent quark mass is $178$ MeV and for set 5 is $372$ MeV.
The constituent quark masses for other parameterizations are in between these two cases.
Therefore, one can consider set 1 and set 5 as limited cases for the predictions of the NJL model.

\begin{table}[th]
\centering%
\begin{tabular}{|c|c|c|c|c|c|c|c|}
\hline
Set&$m_0$[MeV]&$\Lambda$[MeV]&G[GeV$^{-2}$]&$m$[MeV]\\
\hline
1&3.0  & 1464 &  1.61 & 178 \\
2&5.0  & 1097 &  3.07 & 204 \\
3&8.0  & 849  &  5.85 & 245 \\
4&10.0 & 755  &  8.13 & 265 \\
5&15.0 & 645  & 17.2  & 372 \\
\hline\end{tabular}
\label{TableParameters}
\vspace*{8pt}
\caption{Parameters of the NJL model in the proper-time regularization taken from \cite{Inagaki:2015lma}.
}
\end{table}

\begin{figure}[tb]
\centerline{
\includegraphics[width=0.47\textwidth]{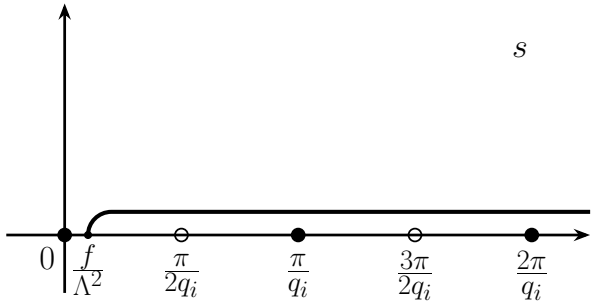}}
\caption{Contour on complex $s$-plane. Singularities for a quark of flavor i which are related to tangent 
are shown by open circles while filled circles correspond to those of cotangent $\cot(q_i s)$.
}
\label{ContTikz}
\end{figure}

The important point of calculation is that integrand of $\mathcal{S}_{eff}$ contain singularities and one should specify how to deal with them. The singularities which are generated by trigonometric functions tangent and cotangent of $q_i s$ for quark flavor i are located at real axis and by hyperbolic functions at imaginary axis.
We shift $s$ to the complex plane $s+i\epsilon$, see Fig. \ref{ContTikz}, since we prefer to running a numerical calculation of integral instead of residues summation like what used in \cite{Inagaki:2003ac,Ruggieri:2016xww}.  
In principle, the effective potential at finite $f$ acquires an imaginary part which correspond to pair-production because of Schwinger mechanism \cite{Schwinger:1951nm,Tavares:2018poq,Cao:2015dya}. 
We figure out that the imaginary part is smaller than the real part in current work. Plus, the subtle effect of Schwinger mechanism is out of the scope of the present paper and will not discuss here.

\begin{figure*}[tb]
\centerline{
\includegraphics[width=0.49\textwidth]{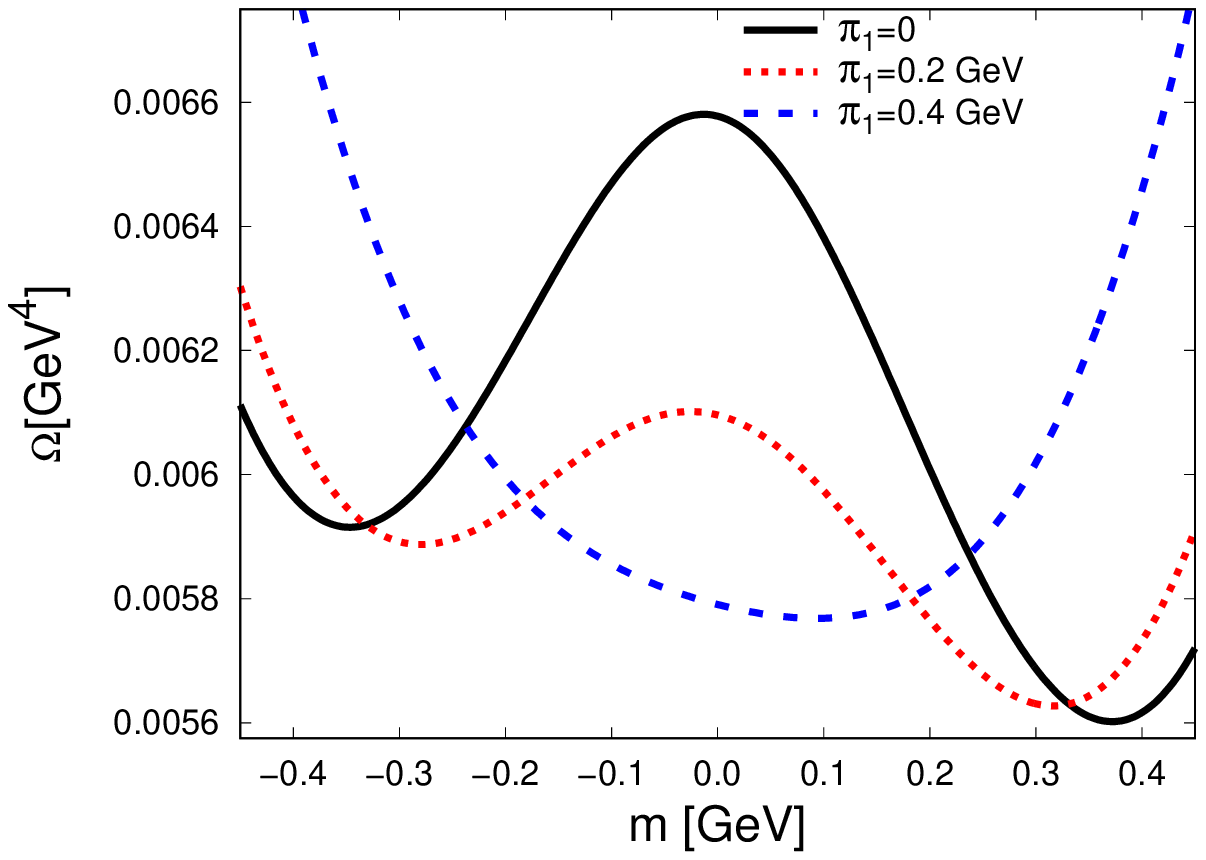}
\includegraphics[width=0.43\textwidth]{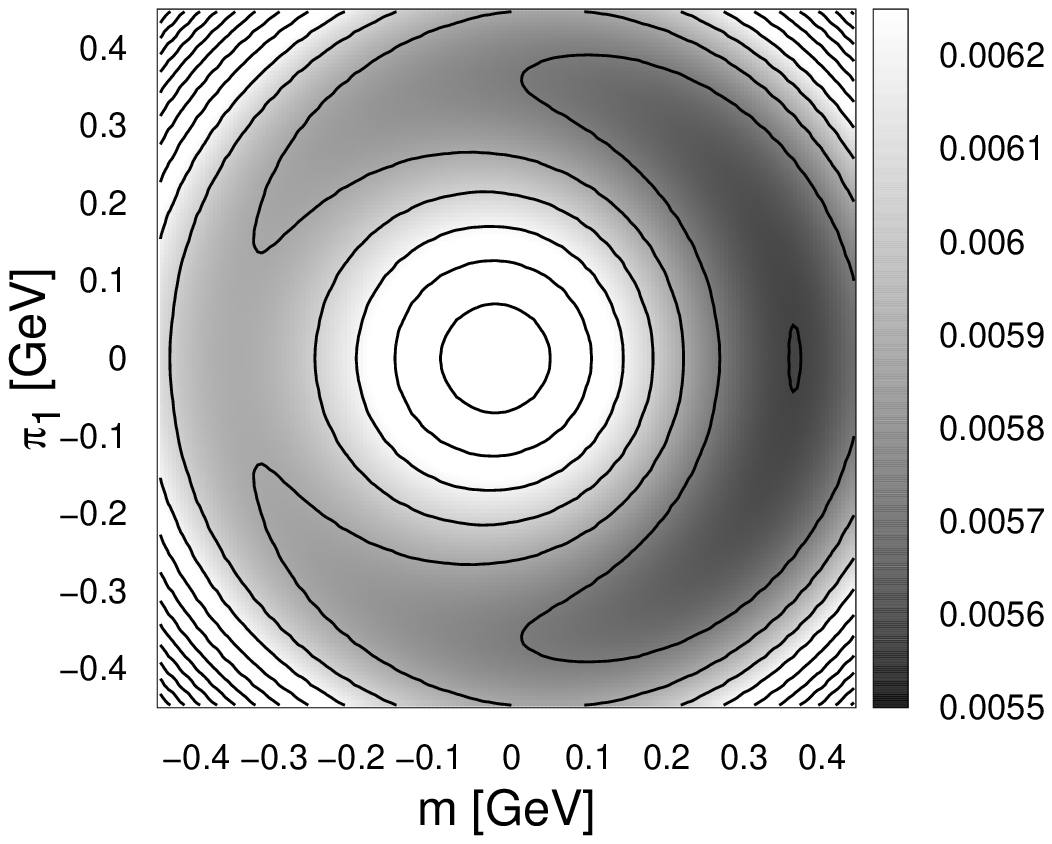}}
\caption{Behavior of effective potential for Set 5 for field value $f=0.01$ GeV$^2$: as a function of quark mass for fixed values of $\pi_1$ (left part), and as a function of mass and  $\pi_1$ (right part). At left part black solid line corresponds to zero $\pi_1$,  red dotted to $\pi_1=0.2$ GeV and blue dashed to $0.4$ GeV.
}
\label{Scan5ff001}
\end{figure*}

\begin{figure*}[tb]
\centerline{
\includegraphics[width=0.49\textwidth]{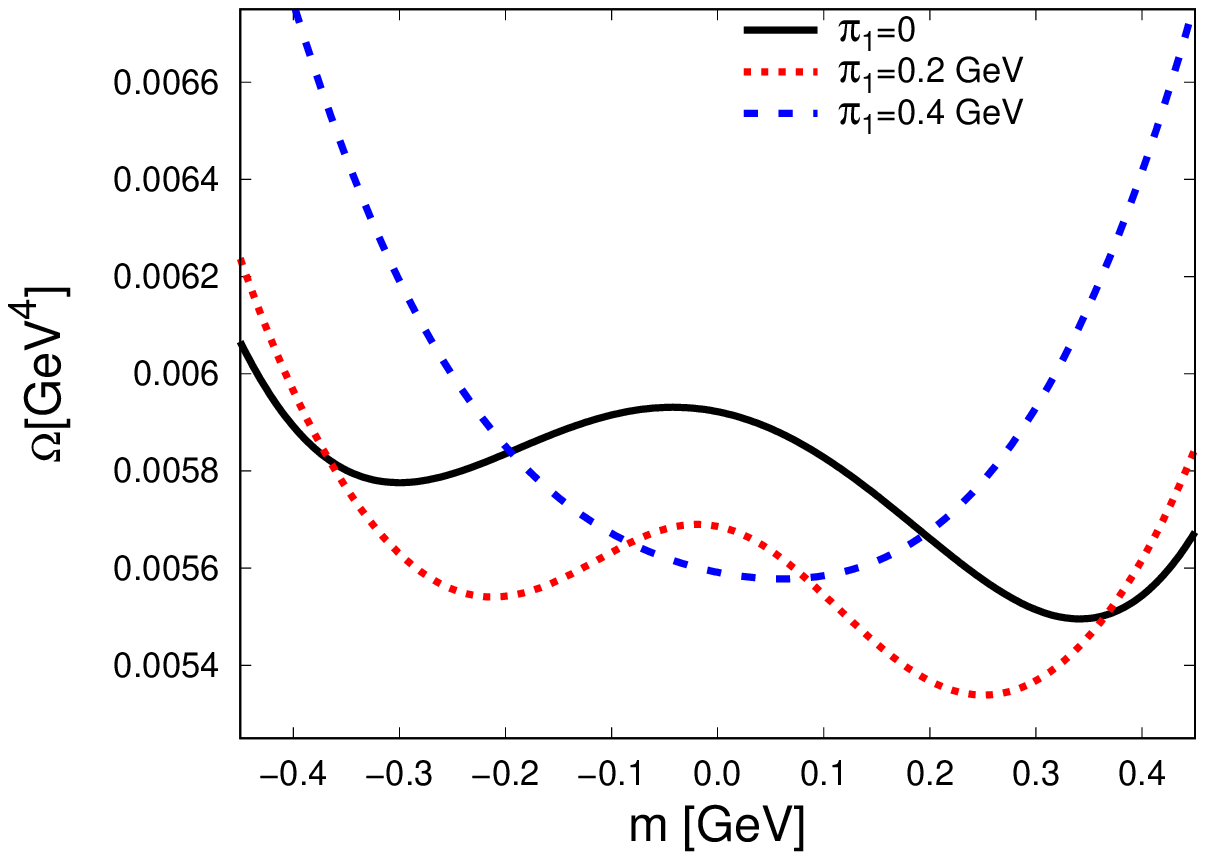}
\includegraphics[width=0.43\textwidth]{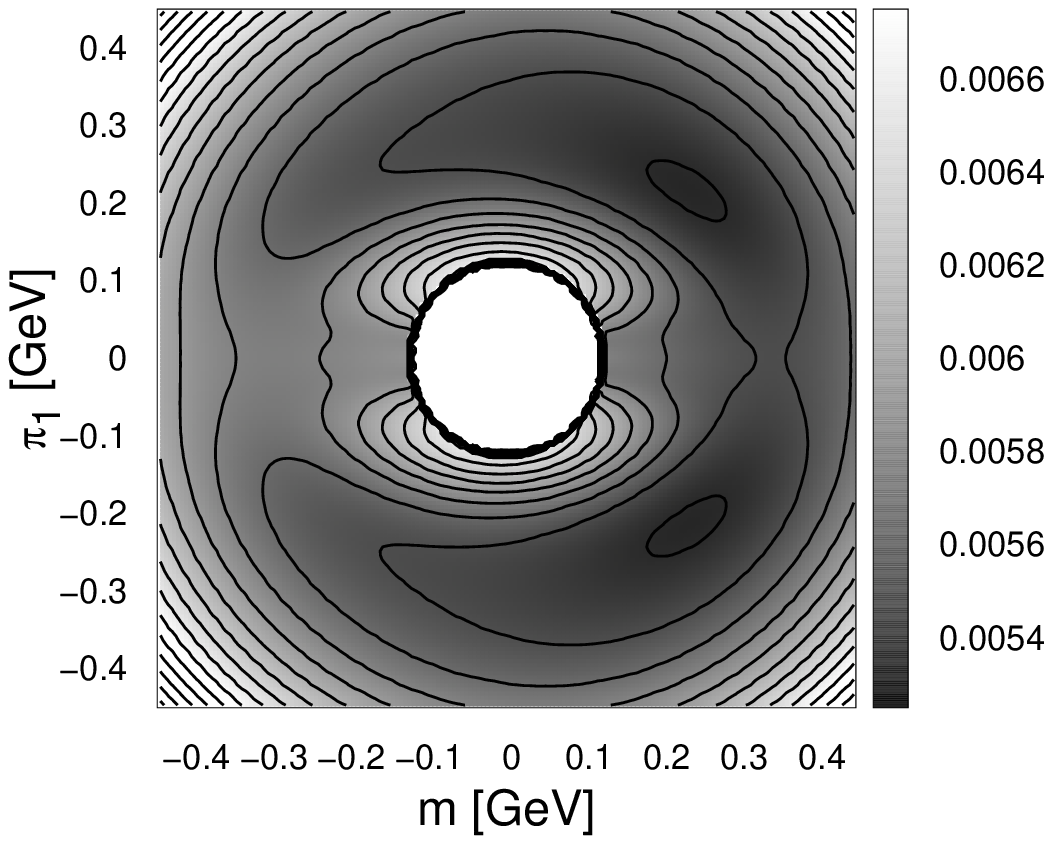}}
\caption{Behavior of effective potential  for Set 5 for field value $f=0.2$ GeV$^2$: as a function of quark mass for fixed values of $\pi_1$ (left part), and as a function of mass and  $\pi_1$ (right part). At left part black solid line corresponds to zero $\pi_1$,  red dotted to $\pi_1=0.2$ GeV and blue dashed to $0.4$ GeV.
The region in center of left part is omitted because when both $m$ and $\pi_1$ are small and nonzero the corrections are nonphysical. 
}
\label{Scan5ff010}
\end{figure*}

\begin{figure*}[tb]
\centerline{
\includegraphics[width=0.49\textwidth]{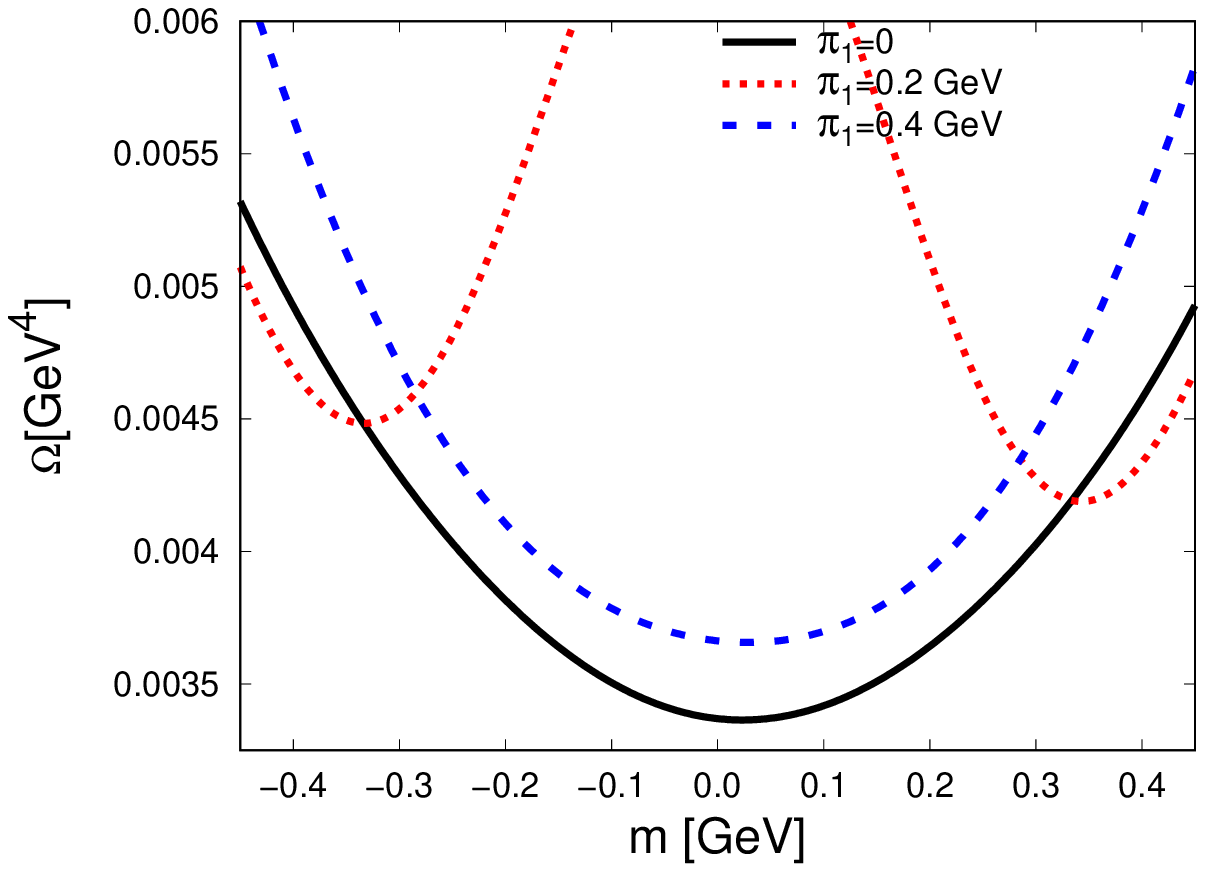}
\includegraphics[width=0.43\textwidth]{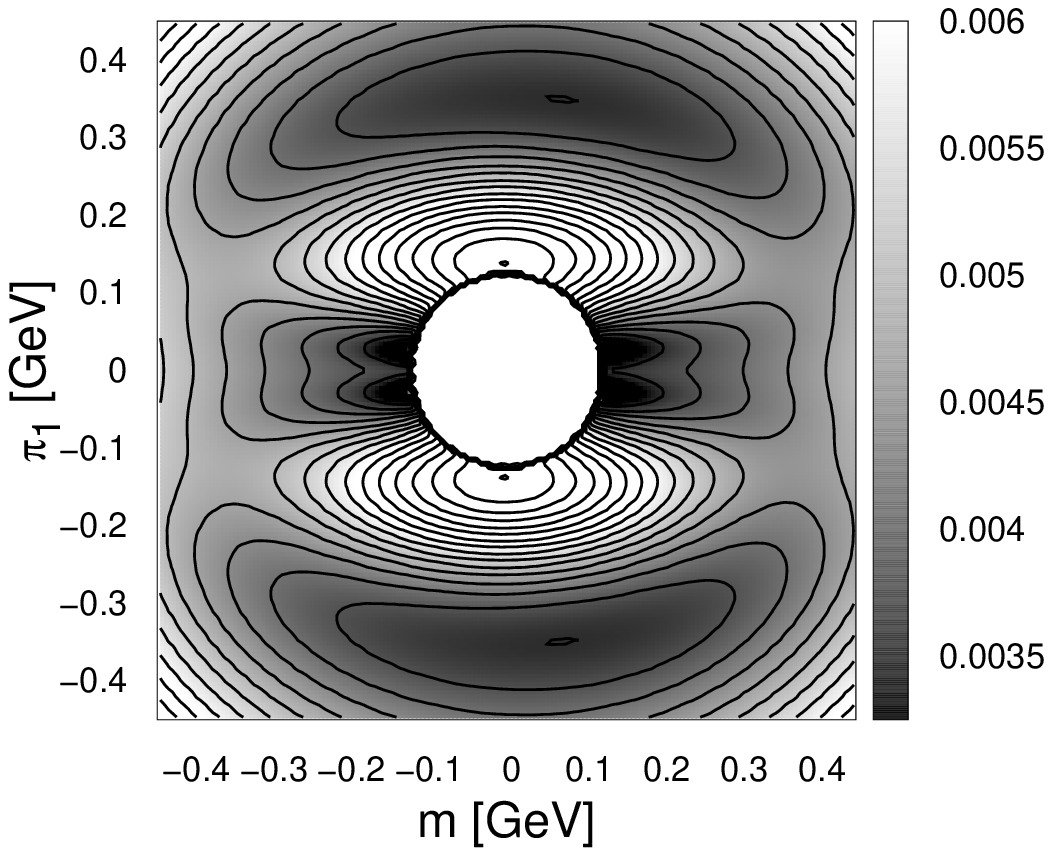}}
\caption{Behavior of effective potential for Set 5 for field value $f=0.45$ GeV$^2$: as a function of quark mass for fixed values of $\pi_1$ (left part), and as a function of mass and  $\pi_1$ (right part). At left part black solid line corresponds to zero $\pi_1$,  red dotted to $\pi_1=0.2$ GeV and blue dashed to $0.4$ GeV.
The region in center of left part is omitted because when both $m$ and $\pi_1$ are small and nonzero the corrections are nonphysical. 
}
\label{Scan5ff035}
\end{figure*}

In Figs. \ref{Scan5ff001}, \ref{Scan5ff010}, \ref{Scan5ff035}, the behavior of effective potential for Set 5 of model parameters is plotted for field values $f=0.01,0.2,0.450$ GeV$^2$, respectively. We found the following typical behavior for three regions: 1) For small field $f=0.01$ GeV$^2$ as shown in Fig. \ref{Scan5ff001}, the system is in usual (almost vacuum) chiral symmetry breaking phase with nonzero sigma condensate and zero pion condensate; 2) For moderate field $f=0.2$ GeV$^2$, seen in Fig. \ref{Scan5ff010}, the additional minima appears in the effective potential 
and the system takes a chiral rotation in $\sigma -\pi_1$ plane to have a nonzero pion condensate, $\pi_1$; 3)  For large field $f=0.450$ GeV$^2$, read from Fig. \ref{Scan5ff035}, the minimum with $\pi_1=0$ is energetically favorable.

There are two sources to break the chiral symmetry: spontaneous chiral symmetry breaking due to presence of quark condensate $\<\bar{\psi} \psi \>$ 
and explicit chiral symmetry breaking due to nonzero current quark mass in the Lagrangian. Therefore, we investigate not only the reality situation but also for $m_0\to 0$. To systematically perform this task, we vary $m_{0}$ and recalculate $m$ while $\mathrm{G}$ and $\Lambda$ have the same values, i.e. we consider $m$ as a function of $m_{0}$ \cite{Bernard:1992mp}. In the following we denote the physical value of current quark mass as $m_{0}^\star$. 

\begin{figure*}[tb]
\centerline{
\includegraphics[width=0.47\textwidth]{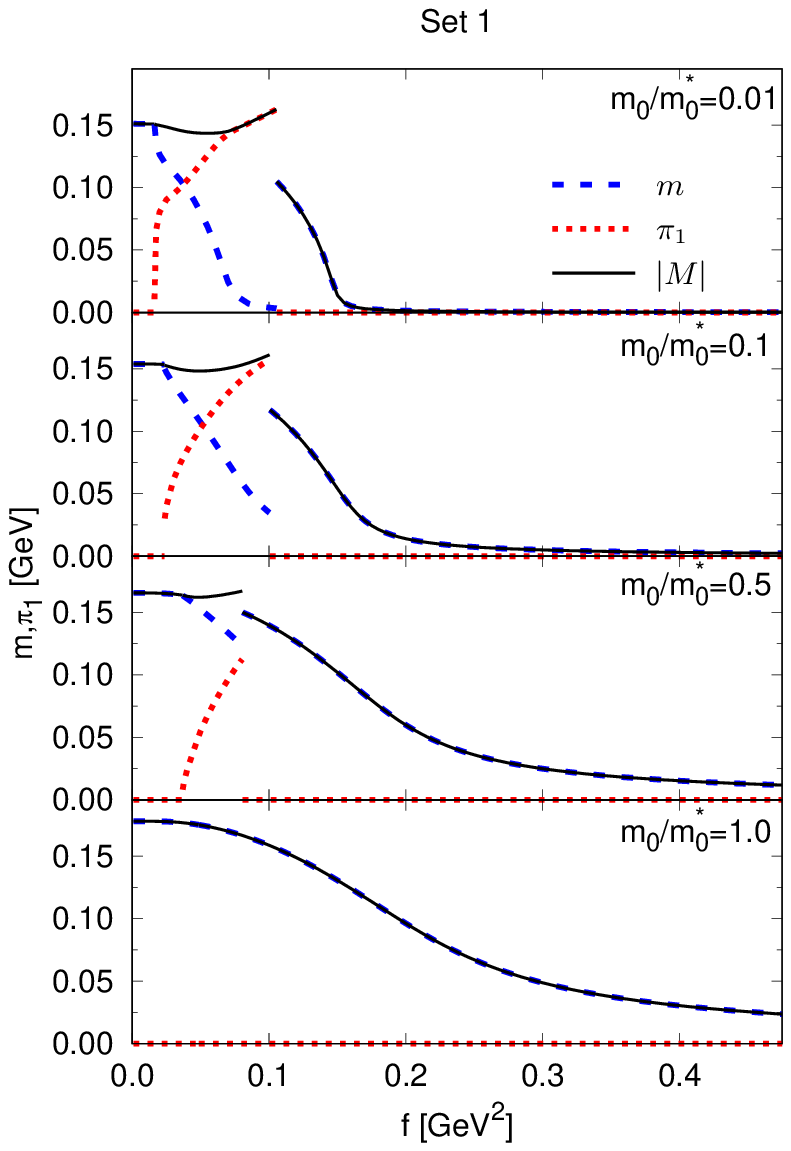}
\includegraphics[width=0.47\textwidth]{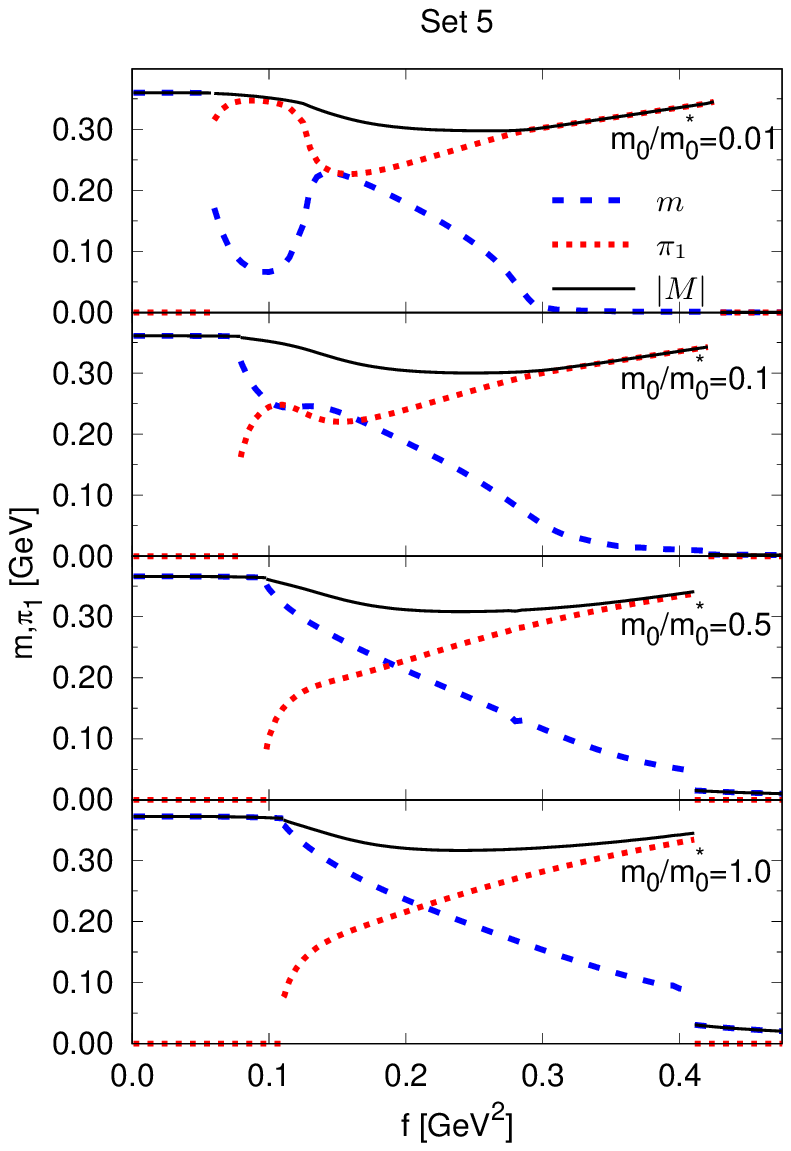}}
\caption{Behavior of quark mass $m$ (blue dashed line), pion condensate $\pi_1$ (red dotted line)  
and their combination $|M|=\sqrt{m^2+\pi_1^2}$ (black solid line) as a function of field $f$ for Sets $1,5$ of model parameters for different values of ratio of current quark mass to its physical value
$m_0/m_0^\star=0.01,0.1,0.5,1.0$ from top 
to bottom. 
}
\label{Set1Set5MassDelta}
\end{figure*}

The behaviors of $m$ and $\pi_1$  as a function of field $f$ are presented in Fig. \ref{Set1Set5MassDelta} for different values of ratio $m_{0}/m_{0}^\star=0.01,0.1,0.5,1.0$. The left and right sides are obtained by model parameter sets 1 and 5, respectively. It is straightforward to figure out that for small current quark mass the system is more preferable to chirally rotate from zero to nonzero value $\pi_1$, leaving the total order parameter of chiral symmetry breaking $|M|=\sqrt{m^2+\pi_1^2}$ unchanged. With increasing of $m_{0}$ the situation becomes more complicated. The phase of pion condensation even never show up for $m_{0}=m_{0}^\star$ in the model parameter Set 1.

\section{Conclusions}
\label{con}
In this paper the charged pion condensation under the parallel electromagnetic fields is calculated in the framework of the NJL model by using Schwinger proper-time method. The configuration of field is chosen, the electric field being anti-parallel to the magnetic one, to have a zero first Lorentz invariant, $I_{1}=\mathbf{E}^{2}-\mathbf{B}^{2}$, and a nonzero second Lorentz invariant, $I_2=\mathbf{E} \cdot \mathbf{B}$.

We find that in the chiral limit the system is favorable to form a both nonzero condensation of scalar and charged pion, i.e. rotating in the chiral group. Chiral condensates aligning to pseudo mesons space has been found in \cite{Cao:2015cka} by the methods of $\chi\mr{PT}$ and NJL model, where the system is immediately straighten up $\pi_0$ direction in the chiral limit once the second Lorentz invariant $I_2$ turned on. The main difference of charged condensation is that the system will across a weakly first order phase transition to zero pion condensate and then a second order phase transition to chirally symmetric phase as the field strength increasing, while it, characterizing by $\pi_{0}$, is a whole second order phase transition as shown in \cite{Cao:2015cka}.
The underlying mechanism are two folds. One is the obviously coupling between charged pions and electromagnetism. Another reason is that a more complicated influence of anomalous diagrams are implicitly included, not only $\pi_0\rightarrow\gamma\gamma$ but also $\gamma\rightarrow \pi_+\pi_-\pi_0$.

Indeed, if assuming condensation in the neutral channel $\<\sigma\>$ nears a second order phase transition, its effective potential has the form $\mc{S}_{eff}^{0}\sim -c_{0}M^{2}+c_{1}M^{4}/f$ according to Ginzburg-Landau theory \cite{Ginzburg:1950sr}. However, if we include $\pi_{\pm}$ as an additional degree of freedom and non-degenerate with $\pi_{0}$, read from \Eqn{eqn_eff_potentioal}, the potential arranges as:
$\mc{S}_{eff}^{2}\sim -\tilde{c}_{1}M^{4}/f+c_{2}M^{6}/f^{2}$. As a result, we have a weakly first order phase transition and effective potential in the form of
\begin{align}
	\Om=\frac{M^{2}}{4 \mathrm{G}}-c_{0}M^{2}+\frac{\l(c_{1}-\tilde{c}_{1}\r)M^{4}}{f}+\frac{c_{2}M^{6}}{f^{2}}.
\end{align}
Our numerical simulations support these arguments, read from Figs. \ref{Scan5ff001}, \ref{Scan5ff010}, \ref{Scan5ff035}. The mass of current quarks plays an important role and it denies our claim at some regions of the model parameters. It requires a further study via the first principle calculation, such as Dyson-Schwinger equation or functional renormalization group methods.

Application of the charged pion condensation to the case heavy-ion collisions or neutron stars interior need an extension to finite temperature and/or chemical potential. We will explore this extension in future.

\section{Acknowledgments}
\label{Ackn}
We are grateful to Maxim Chernodub, Nikolai Kochelev, Marco Ruggieri and Pengming Zhang for the useful discussions.
J.Y.C. is supported by the NSFC under Grant number: 11605254 and Major State Basic Research Development Program in China (No. 2015CB856903).
M.H. is supported by the NSFC under Grant No. 11725523, 11735007 and 11261130311(CRC 110 by DFG and NSFC).
A.R. is supported by the CAS President's international fellowship initiative (Grant No. 2017VMA0045), 
Council for Grants of the President of the Russian Federation (project NSh-8081.2016.9) and
 numerical calculations are performed on computing cluster "Akademik V.M. Matrosov" (http://hpc.icc.ru).



\end{document}